\documentclass{lmcs}
\pdfoutput=1

\usepackage{lastpage}
\renewcommand{\dOi}{14(4:3)2018}
\lmcsheading{}{1--\pageref{LastPage}}{}{}%
{Feb.~21,~2017}{Oct.~24,~2018}{}

\usepackage{hyperref}
\usepackage{amssymb, amsmath}
\usepackage{bm}
\usepackage{tikz}
\usetikzlibrary{arrows}

\newcommand{\true}{1}
\newcommand{\false}{0}
\newcommand{\semblocked}{\mathsf{SEM_{BC}}}

\newcommand{\rat}{\mathsf{RAT}}

\newcommand{\env}[1]{\mathit{env}_{#1}}
\newcommand{\envf}{\env{F}}
\newcommand{\glob}[1]{\mathit{ext}_{#1}}

\newcommand{\var}{\mathit{var}}
\newcommand{\sat}{\textsc{SAT}}
\newcommand{\p}{\textsf{P}}
\newcommand{\np}{$\mathsf{NP}$}
\newcommand{\conp}{$\mathrm{co}$-\np}
\newcommand{\pitwo}{$\Pi^{P}_2$}
\newcommand{\aesat}{$\forall\exists$-\sat}
\newcommand{\ctrans}[1]{t}
\newcommand{\ctransf}{\ctrans{F}}
\newcommand{\rtaut}{\mathsf{RT}}
\newcommand{\rsubs}{\mathsf{RS}}
\newcommand{\rasubs}{\mathsf{RAS}}
\newcommand{\taut}{\mathsf{T}}
\newcommand{\subs}{\mathsf{S}}
\newcommand{\bc}{\mathsf{BC}}
\newcommand{\ataut}{\mathsf{AT}}
\newcommand{\asubs}{\mathsf{AS}}
\newcommand{\abc}{\mathsf{ABC}}
\newcommand{\reso}{\mathsf{R}}
\newcommand{\setb}{\mathsf{SET_{BC}}}
\newcommand{\superb}{\mathsf{SUP_{BC}}}
\newcommand{\rup}{\mathsf{RUP}}
\newcommand{\figvarb}[1]{$\bm{#1}$}
\newcommand{\figvar}[1]{$#1$}
\newcommand{\posints}{\mathbb{N}^{+}}
\newcommand{\tauliterals}{T}
\newcommand{\res}{\otimes}
\newcommand{\quant}{\mathcal{Q}}

\begin{document}

\title[Local Redundancy in SAT]{Local Redundancy in SAT:\texorpdfstring{\\}{} Generalizations of Blocked Clauses}

\author[]{Benjamin Kiesl\rsuper{1}}	
\address{\lsuper{1}Institute of Logic and Computation, TU Wien}
\email{kiesl@kr.tuwien.ac.at}  
\thanks{This work has been supported by the Austrian Science Fund (FWF) under projects W1255-N23 and S11408-N23.}	
\address{\vskip-7pt}	
\email{tompits@kr.tuwien.ac.at}  

\author[]{Martina Seidl\rsuper{2}}	
\address{\lsuper{2}Institute for Formal Models and Verification\\
Johannes Kepler University}
\email{martina.seidl@jku.at}  

\author[]{Hans Tompits\rsuper{1}}	
\address{\vskip-7pt}

\author[]{Armin Biere\rsuper{2}}	
\address{\vskip-7pt}	
\email{biere@jku.at}  


\subjclass{Theory of Computation, Logic, Automated Reasoning}
\titlecomment{
This is an extended version of the paper ``Super-Blocked Clauses''~\cite{superblocked_clauses}
which has appeared in the Proceedings of the 8th International Joint Conference on Automated Reasoning (IJCAR~2016)}


\begin{abstract}
\noindent
Clause-elimination procedures that simplify formulas in conjunctive normal form play
an important role in modern SAT solving.
Before or during the actual solving process, such procedures identify and remove
clauses that are irrelevant to the solving result.
These simplifications
usually rely on so-called redundancy properties that characterize
cases 
in which the removal of a clause 
does not affect the satisfiability status of a formula.
One particularly successful redundancy property is that of
blocked clauses, because it generalizes
several other redundancy properties.
To find out whether a clause is blocked---and therefore redundant---one only needs
to consider its resolution environment, i.e., 
the clauses with which it can be resolved. 
For this reason, we say that the redundancy property of blocked clauses 
is local.
In this paper, we show that there exist
local redundancy properties that are even more general than blocked clauses.
We present a semantic notion of blocking and prove that it constitutes the
most general local redundancy property. 
We furthermore introduce the syntax-based notions of
set-blocking and super-blocking, and show that the latter coincides with our semantic blocking notion.
In addition,
we show how semantic blocking can be alternatively characterized via 
Davis and Putnam's rule for eliminating atomic formulas.
Finally, we perform a detailed complexity analysis and 
relate our novel redundancy properties to prominent redundancy properties
from the~literature.
\end{abstract}

\maketitle

\section*{Introduction}\label{sec:intro}

Over the last two decades, there has been enormous progress in the 
performance of SAT solvers, i.e., decision procedures for the 
satisfiability problem of propositional logic (SAT)~\cite{DBLP:series/faia/2009-185}. 
As a consequence, 
SAT solvers have become attractive reasoning engines in many user 
domains like the verification of 
hardware and software~\cite{vizel15} as well as in the backends 
of other reasoning tools like SMT 
solvers~\cite{DBLP:series/faia/BarrettSST09},
QBF solvers~\cite{janota16_qbfcegar, samulowitz05_usingsatinqbf},
or even first-order theorem provers~\cite{DBLP:conf/cade/RegerSV15}. 
In such applications, however, SAT solvers often reach their limits,
motivating the quest for more efficient SAT techniques.

One successful approach to improving the performance of SAT solvers is the use
of clause-elimination procedures.  
Either before (``preprocessing'') 
or during the actual solving (``inprocessing'') such procedures simplify
a given formula in conjunctive normal 
form (CNF)
by removing clauses that are irrelevant to the outcome of the solving process~\cite{OptimizedBlockedClauseDecomposition,
DBLP:journals/corr/Chen15o,
BlockedClauseDecomposition,
ClauseEliminationJournal,
DBLP:conf/sat/IserMS15,
BlockedClauseEliminationJournal,
Inprocessing,
DBLP:conf/sat/MantheyPW13}.
To distinguish the irrelevant from the relevant, they usually rely on
so-called
\emph{redundancy properties} that 
characterize cases in which
certain clauses can be removed from a formula without affecting its satisfiability status~\cite{ClauseEliminationJournal,Inprocessing}.
For instance, a clause-elimination procedure that
is based on the simple redundancy property of \emph{tautologies} identifies and removes 
tautological clauses---clauses that contain two complementary literals.
As these clauses are true regardless of a particular truth assignment, their
removal does not influence the outcome of the solving process.

A particularly useful redundancy property 
is that of \emph{blo\-cked clauses}~\cite{BlockedClauseElimination,BlockedClauses}.
Informally, a clause $C$ is \emph{blocked} in a CNF formula $F$ if it contains a literal~$l$ such that 
all possible resolvents of $C$ upon $l$ (with clauses from $F$) are tautologies.
The elimination of blocked clauses considerably improves the performance of modern SAT solvers~\cite{BlockedClauseElimination, DBLP:conf/sat/MantheyPW13}.
Moreover, blocked clauses provide the basis for \emph{blocked-clause decomposition}, a technique that 
splits a CNF formula into two parts that become solvable by 
blocked-clause elimination~\cite{BlockedClauseDecomposition}.
Blocked-clause decomposition is successfully used for gate extraction, for
efficiently finding backbone variables, and for the detection of 
implied binary equivalences~\cite{OptimizedBlockedClauseDecomposition,DBLP:conf/sat/IserMS15}.
The winner of the SAT-Race 2015 competition, the solver 
$\mathsf{abcdSAT}$~\cite{DBLP:journals/corr/Chen15o},
uses blocked-clause decomposition as core technology. 

Part of the reason for the success of blocked clauses is the fact that they
generalize several other well-known redundancy properties such as those of \emph{pure literals}~\cite{davis60_a_computing_procedure} or the above-mentioned tautologies. 
This means that a tool for blocked-clause elimination implicitly performs the elimination of pure literals and  
tautologies (and 
even more aggressive formula simplifications 
on top).
A closer look at the definition of blocked clauses reveals that to
check whether a clause is blocked, it suffices to consider only its resolution environment, i.e., 
the clauses with which it can be resolved.
Because of this, we say that the blocked-clauses redundancy property is \emph{local}.
Their locality aspect makes blocked clauses particularly effective when
dealing with large formulas in which the resolution environments of clauses
are~small.

These success stories motivate us to have 
a closer look at local redundancy properties in general and at blocked clauses in particular.
We show in this paper that blocked clauses do not constitute the most general concept of a local redundancy property.
In particular, we first present some observations on blocked clauses and provide
examples of redundant clauses that are not blocked, yet their redundancy can be identified
by considering only their resolution environment. 
We next use these observations to derive a 
semantic notion of blocking for which we show that it constitutes the most general local redundancy property.
To bring this semantic notion of blocking closer to practical SAT solving, we then introduce the syntax-based
redundancy properties of \emph{set-blocked clauses} and \emph{super-blocked clauses}---both are strictly more general
than traditional blocked clauses and for super-blocking we prove that it coincides with our semantic blocking notion.

After introducing the new redundancy properties, we present an alternative 
syntactic characterization of semantic blocking based
on Davis and Putnam's \emph{rule for eliminating atomic formulas}~\cite{davis60_a_computing_procedure}---a 
resolution-based rewriting rule that is also known as
\emph{variable elimination}~\cite{effective_preprocessing_een05} in the context of SAT solving.
This characterization further clarifies the connection between traditional blocking and semantic blocking.
We then proceed by analyzing the complexity of deciding whether a clause is redundant with respect
to our novel redundancy properties. Our complexity analysis gives rise to a whole family of redundancy properties
that are obtained by restricting the notions of set-blocking and super-blocking in certain ways.
Finally, we outline the relationship of the new redundancy properties to prominent redundancy properties
from the literature before concluding
with an outlook to future work.

A preliminary version of this paper appeared in the proceedings of IJCAR 2016~\cite{superblocked_clauses}.
Besides several extensions of the text, 
this version now includes the full proofs of our complexity results while the 
conference paper~\cite{superblocked_clauses} contained only proof sketches. In addition, we introduce an alternative characterization of 
semantic blocking based on variable elimination~(cf.\ Section~\ref{sec:ve}).

\section{Preliminaries}
\label{sec:prelim}

We consider propositional formulas in \emph{conjunctive normal form} (CNF) which 
are defined as follows. 
A \emph{literal} is either a Boolean variable $x$ (a \emph{positive literal}) or 
the negation $\neg x$ of a variable $x$ (a \emph{negative literal}). 
For a literal $l$, we define $\bar l = \neg x$ if $l = x$ and $\bar l = x$ if $l = \neg x$.
Accordingly, for a set $L$ of literals, we define $\bar{L} = \{\bar l \mid l \in L\}$. 
A \emph{clause} is a disjunction of literals. A \emph{formula} 
is a conjunction of clauses.
We identify a clause with a set of literals and a formula with a set of clauses.
A \emph{tautology} is a clause that contains both $l$ and $\bar{l}$ for some literal $l$.
For a literal, clause, or formula $E$, $\var(E)$ denotes the set of variables occurring in $E$. 
For convenience, we often treat $\var(E)$ as a variable if $E$ is a literal.
For a set $L$ of literals and a formula $F$, we denote by $F_L$ the set of all clauses in $F$ that contain a literal of $L$, 
i.e., $F_L = \{C \mid \text{$C \in F$ and $C \cap L \neq \emptyset$}\}$.
In case $L$ is a singleton set of the form $\{l\}$, we sometimes write $F_l$ instead of $F_{\{l\}}$.

An \emph{assignment} is a 
function from a set of variables to the truth values 
\true{}~(\emph{true}) and \false{} (\emph{false}).
An assignment is \emph{total} with respect to a formula if it assigns a truth value to all
variables occurring in the formula.
Unless stated otherwise, we do not assume assignments to be total. 
Assignments are extended from variables to literals by defining $\tau(l) = \tau(\var(l))$ if the literal $l$ is positive and by defining $\tau(l) = 1 - \tau(\var(l))$ if $l$ is negative.
Given an assignment $\tau$ that assigns a truth value to a literal~$l$, we denote by $\tau_l$ the assignment obtained from $\tau$ by interchanging (``flipping'') 
the truth value of $l$, i.e., by defining  $\tau_l(x) = 1 - \tau(x)$ if $x = \var(l)$ and \mbox{$\tau_l(x) = \tau(x)$} otherwise.

A literal $l$ is \emph{satisfied} by an assignment $\tau$ if $\tau(l) = \true$.
A clause is satisfied by an assignment $\tau$ if it contains a literal that is satisfied by $\tau$.
Finally, a formula is satisfied by an assignment $\tau$ if all of its clauses are satisfied by $\tau$.
A formula is \emph{satisfiable} if there exists an assignment that satisfies it.
Two formulas are \emph{logically equivalent} if they are satisfied by the same total assignments.
Two formulas $F$ and $F'$ are \emph{satisfiability equivalent} if 
either both $F$ and $F'$ are satisfiable or both $F$ and $F'$ are unsatisfiable. 
We sometimes say that an assignment $\tau$ \emph{falsifies} a literal~$l$ if $\tau(l) = \false$.
Accordingly
$\tau$ falsifies a clause $C$ if $\tau$ falsifies all literals of $C$.

Given two clauses $C$ and $D$ together
with a literal $l \in C$ such that $\bar{l} \in D$, 
the clause $(C \setminus \{l\}) \cup (D \setminus \{\bar{l}\})$ is 
the \emph{resolvent} of $C$ and $D$ upon $l$ and denoted by $C \res_{l} D$.
Given a formula $F$ and a clause $C$, the \emph{resolution environment} 
of $C$ in $F$ is the set of all clauses in $F$ that can be resolved with $C$:

\begin{defi}
The \emph{resolution environment} of a clause $C$ with respect to a formula $F$ is the clause set
$\envf(C) = \{D \in F \mid \exists \,l \in D \text{ such that } \bar{l} \in C\}$.
\end{defi}

\noindent
Given a formula $F$ and a clause $C$, we call the variables in $\var(C)$ the \emph{local variables} and the variables in $\var(\envf(C)) \setminus \var(C)$ the $\emph{external variables}$. We denote the set of external variables by $\glob{F}(C)$.

Next, we formally introduce redundancy of clauses. 
Intuitively, a clause $C$ is redundant with respect to a formula $F$ if 
neither its addition to $F$ nor its removal from $F$ changes the satisfiability or unsatisfiability of $F$.

\begin{defi}
A clause $C$ is \emph{redundant} with respect to a formula $F$ if $F \setminus \{C\}$ and $F \cup \{C\}$ are satisfiability equivalent.
A \emph{redundancy property} is a set of pairs $(F,C)$ where $C$ is redundant with respect to $F$.
Finally, for two redundancy properties $\mathcal{P}_1$ and $\mathcal{P}_2$,
$\mathcal{P}_1$ is \emph{more general} than $\mathcal{P}_2$ if $\mathcal{P}_2 \subseteq \mathcal{P}_1$,
i.e., if every pair $(F,C)  \in \mathcal{P}_2$ is also contained in $\mathcal{P}_1$.
If $\mathcal{P}_2 \subset \mathcal{P}_1$, then $\mathcal{P}_1$ is \emph{strictly more general} than $\mathcal{P}_2$. 
\end{defi}

\begin{exa}
Consider the formula $F = \{ (a \lor b), (\neg a \lor \neg b) \}$. 
The clause \mbox{$C = (\neg a \lor \neg b)$} is redundant with respect to $F$ since $F \setminus \{C\}$ and $F \cup \{C\}$ are satisfiability equivalent 
\textup{(}\kern-.15em although they are not logically equivalent\textup{)}.
Moreover, the set \[\{(F,C) \mid \text{$F$ is a formula and $C$ is a tautology}\}\] is a redundancy property since 
for every formula $F$ and every tautology~$C$, 
$F \setminus \{C\}$ and $F \cup \{C\}$ are satisfiability equivalent.
\end{exa}

\noindent
Note that a clause $C$ is \emph{not} redundant with respect to a formula $F$ if and only if
$F \setminus \{C\}$ is satisfiable and $F \cup \{ C \}$ is 
unsatisfiable. To prove that $C$ is redundant with respect to $F$, it therefore
suffices to show that satisfiability of $F \setminus \{C\}$ implies satisfiability of $F \cup \{C\}$.
Redundancy properties as defined above yield
the basis not only for clause-elimination but also for 
clause-addition procedures~\cite{Inprocessing}.

\section{Observations on Blocked Clauses}
\label{sec:observations}

Following Heule et al.~\cite{ClauseEliminationJournal}
we recapitulate the definition of blocked clauses.
In the rest of the paper, we refer 
to this kind of blocked clauses as \emph{literal-blocked clauses}. 
Motivated by
the examples given in this section, we then
generalize this traditional notion of blocking to more powerful redundancy properties.

\begin{defi}\label{def:bc_original}
	A clause $C$ is \emph{blocked} by a literal $l \in C$ in a formula $F$ if
	   $C \cup (D \setminus \{\bar{l}\})$ 
	  is a tautology 
	for each clause $D \in F_{\bar l}$.
	A clause $C$ is \emph{literal-blocked} in $F$ if there exists a literal that blocks $C$ in $F$.
	By $\bc$ we denote the set $\{(F,C) \mid \text{$C$ is literal-blocked in $F$}\}$.
\end{defi}

\begin{exa}\label{ex:blocked}
Consider the formula $F = \{(\neg a \lor c), (\neg b \lor \neg a)\}$ and the clause $C = (a \lor b)$. 
The literal $b$ blocks $C$ in $F$ since the only clause in $F$ that contains $\neg b$ is the clause $D$ 
with
$D = (\neg b \lor \neg a)$ and further $C \cup (D \setminus \{\bar{l}\}) = (a \lor b \lor \neg a)$ is a tautology. 
\end{exa}

\noindent
Note that this definition of literal-blocked clauses differs slightly from Kullman's original definition~\cite{BlockedClauses}. The original definition requires that all \emph{resolvents} of $C$ upon a literal $l$ are tautologies. However, it is not necessary to remove $l$ from $C \cup (D \setminus \{\bar{l}\})$---as would be the case in the resolvent of $C$ and $D$---to guarantee redundancy of literal-blocked clauses. If $l$ is not removed from $C$, every tautology $C$ is also a literal-blocked cause, which is not the case for the original definition.

\begin{prop}
\label{thm:blocked_clauses_are_redundant}
$\bc$ is a redundancy property.
\end{prop}

\noindent Proposition~\ref{thm:blocked_clauses_are_redundant} paraphrases results from Heule et al.~\cite{ClauseEliminationJournal} 
and actually follows from our present results given later on 
(namely from Proposition \ref{thm:bc_subset_setb} and Corollary \ref{thm:setb_is_a_local_redundancy_property}).
To see that blocked clauses are redundant, let $C$ be a clause that is blocked by a literal $l$ in
a formula $F$ and assume that some assignment $\tau$ satisfies $F \setminus \{C\}$ but
falsifies $C$.
We can then easily turn $\tau$ into a satisfying assignment $\tau_l$ of 
$C$ by flipping the truth value of $l$.

This flipping could only possibly falsify some of the clauses in $F \setminus \{C\}$ that contain $\bar l$, 
but the condition that $l$ blocks $C$ guarantees that these clauses stay satisfied:
Let $D$ be such a clause that contains $\bar l$. 
Then, since $C \cup (D \setminus \{\bar l\})$
is a tautology, either $D$ is itself a tautology (and therefore trivially satisfied) or it contains
a literal $l' \neq l$ such that $\bar l' \in C$.
In the latter case, since $\tau$ falsifies $C$, it must satisfy $l'$ and
since $\tau_l$ agrees with $\tau$ on all literals but $l$,
$\tau_l$ is a satisfying assignment of $F \cup \{C\}$.
Hence, $C$ is redundant with respect to $F$.

Next, we illustrate with an example how a satisfying assignment
of $F \cup \{C\}$ can be obtained from one of $F \setminus \{C\}$~\cite{ClauseEliminationJournal}. 
This approach is used to obtain a satisfying assignment of the original formula when literal-blocked clauses have been removed 
during pre- or inprocessing:
Suppose a SAT solver gets an input formula $F$ and removes literal-blocked clauses to obtain a simplified formula $F'$.
The solver then proceeds by searching for a satisfying assignment of $F'$.
Once it has found such an assignment, it can easily turn it into a satisfying assignment of the original formula~$F$.

\begin{exa}
Consider again the formula $F = \{(\neg a \lor c), (\neg b \lor \neg a)\}$ and the clause $C = (a \lor b)$ from Example \ref{ex:blocked}. 
We already know that $b$ blocks $C$ in $F$. 
So, let $\tau$ denote the assignment that falsifies the variables $a$, $b$, and $c$.
Clearly, $\tau$ satisfies $F$ but falsifies $C$.
Now, the assignment $\tau_b$, obtained from $\tau$ by flipping the truth value of $b$, satisfies
not only $C$ but also all clauses of $F$:
The only clause that could have been falsified by flipping the truth value of $b$ is $(\neg b \lor \neg a)$. 
But, since $\tau$ satisfies $\neg a$ and $\tau_b$ agrees with $\tau$ on all variables except $b$,
$\tau_b$ satisfies $F \cup \{C\}$.
\end{exa}

\noindent
Literal-blocked clauses generalize many other redundancy properties like 
\emph{pure literal} or \emph{tautology}~\cite{ClauseEliminationJournal}.
One of their particularly important properties is that for testing whether some clause $C$ is literal-blocked in a formula $F$ 
it suffices to consider only those clauses of $F$ that can be resolved with $C$, i.e., 
the clauses in the resolution environment $\envf(C)$ of~$C$. 

This raises the question whether there exist redundant clauses that are not blocked
but whose redundancy 
can be identified by considering only their resolution environment.
As shown in the next example, this is indeed the case:

\begin{figure}[b]
\begin{center}
	\begin{tikzpicture}[auto]
		\tikzstyle{complementline}=[solid,line width=0.07mm]
	
		\node (c) at (0,0) {\figvarb{a} $\lor$ \figvarb{b}};
		\node (c1) at (-2.2,0) {\figvar{x} $\lor$ \figvarb{b} $\lor$ $\neg$\hspace{1pt}\figvarb{a}};
		\node (c2) at (2,0.3) {$\neg$\hspace{1pt}\figvarb{b} $\lor$ $\neg$\hspace{1pt}\figvar{x}};
		\node (c3) at (1.889,-0.3) {$\neg$\hspace{1pt}\figvarb{b} $\lor$ \figvarb{a}};
		
		\draw[complementline] (-0.55,0) -- (-1.10,0);
		\draw[complementline] (0.53,0.1) -- (1.2,0.26);
		\draw[complementline] (0.53,-0.1) -- (1.2,-0.26);
		
	\end{tikzpicture}
\end{center}
\vspace{-11pt}
\caption{The clause $(a \lor b)$ from Example \ref{ex:full_blocking} and its resolution environment.}
\label{fig:full_blocking}
\vspace*{-0.1cm}
\end{figure}

\begin{exa}\label{ex:full_blocking}
Let $C  = (a \lor b)$ and $F$ an arbitrary formula with the resolution environment 
$\envf(C) = \{(x \lor b \lor \neg a), (\neg b \lor \neg x), (\neg b \lor a)\}$ \textup{(}cf.\ Fig.\ \ref{fig:full_blocking}\textup{)}.
The clause $C$ is not literal-blocked in $F$ but redundant: 
Suppose there exists an assignment $\tau$ that satisfies $F$ but falsifies $C$. 
Then, $\tau$ must satisfy either $x$ or $\neg x$. If $\tau(x) = 1$, then $C$ can 
be satisfied by flipping the truth value of $a$, resulting in assignment $\tau' = \tau_a$. 
Since $\tau'(x) = 1$, the clause 
$(x \lor b \lor \neg a)$ stays satisfied. 
In contrast, if $\tau(x) = 0$, we can satisfy $C$ 
by the assignment $\tau''$ obtained from $\tau$ by flipping the truth values of both $a$ and $b$: 
The fact that $\tau''(b) = 1$ guarantees that $(x \lor b \lor \neg a)$
stays satisfied whereas $\tau''(x) = 0$ and $\tau''(a) = 1$ 
guarantee that both $(\neg b \lor \neg x)$ and $(\neg b \lor a)$ 
stay satisfied. 
Since flipping the truth values of literals in $C$ does not affect the truth of clauses 
outside the resolution environment $\env{F}(C)$ of $C$, we obtain in both cases a satisfying assignment of $F$.
\end{exa}

\section{A Semantic Notion of Blocking}
\label{sec:semblock}

In the examples of the preceding section, when arguing that a clause $C$ is redundant with respect to some formula $F$, 
we showed that every assignment $\tau$ that satisfies $F \setminus \{C\}$ but falsifies $C$ 
can be turned into a satisfying assignment $\tau'$ of $F \cup \{C\}$ by flipping the truth values of certain literals in $C$. 
Since this flipping only affects the truth of clauses in $\env{F}(C)$,
it suffices to make sure that $\tau'$
satisfies $\env{F}(C)$ in order to guarantee that it satisfies $F \cup \{C\}$. 
This naturally leads to the following semantic notion of blocking:

\begin{defi}\label{def:semblocked}
A clause $C$ is \emph{semantically blocked} in a formula $F$ if, for every satisfying assignment $\tau$ of $\envf(C)$, 
there exists a satisfying assignment $\tau'$ of $\envf(C)  \cup \{C\}$ such that $\tau(v) = \tau'(v)$ for all $v \notin \var(C)$.
We denote the set $\{(F,C) \mid \text{$C$ is semantically blocked in $F$}\}$ by $\semblocked$.
\end{defi}

\noindent Note that clause $C$ in Example \ref{ex:full_blocking} is semantically blocked in $F$. 
Note also that a clause is semantically blocked if its resolution environment is unsatisfiable.

\begin{thm}\label{thm:semblocked_is_a_redundancy_property}
$\semblocked$ is a redundancy property.
\end{thm}

\proof
Let $F$ be a formula and $C$ a clause that is semantically blocked in $F$. 
We show that $F \cup \{C\}$ is satisfiable if $F \setminus \{C\}$ is satisfiable.
Suppose there exists a satisfying assignment $\tau$ of $F \setminus \{C\}$.
We proceed by a case distinction.

\medskip\noindent
{\sc Case 1:} $C$ contains a literal $l$ with $\var(l) \notin \var(F \setminus \{C\})$.
Then, $\tau$ can be easily extended to a satisfying assignment $\tau'$ of $F \cup \{C\}$ by defining that $\tau'$ satisfies $l$.

\medskip\noindent
{\sc Case 2:} $\var(C) \subseteq \var(F \setminus \{C\})$. 
	In this case, $\tau$ is a total assignment with respect to $F \cup \{C\}$. 
	Suppose that $\tau$ falsifies $C$. 
	It follows that $C$ is not a tautology and so it does not contain a literal $l$ such that $\bar{l} \in C$, hence $C \notin \envf(C)$. 
	Thus, $\envf(C) \subseteq F \setminus \{C\}$ and so $\tau$ satisfies $\envf(C)$. 
	Since $C$ is semantically blocked in $F$, 
	there exists a satisfying assignment $\tau'$ of $\envf(C) \cup \{C\}$ such that $\tau(v) = \tau'(v)$ for all $v \notin \var(C)$. 
	Now, since $\tau'(v)$ differs from $\tau$ only on variables in $\var(C)$, the only clauses in $F$ 
	that could possibly be falsified by $\tau'$ are those with a literal $\bar{l}$ such that $l \in C$. 
	But those are exactly the clauses in $\envf(C)$, so $\tau'$ satisfies $F \cup \{C\}$.

\medskip\noindent
Hence, $C$ is redundant with respect to $F$ and thus $\semblocked$ is a redundancy property.
\qed

\noindent If a clause $C$ is redundant with respect to some formula $F$ and if this redundancy can be 
identified by considering only its resolution environment in $F$, 
then we expect $C$ to be redundant with respect to every formula $F'$ in which $C$ has the same resolution environment as in $F$. 
This leads us to the notion of \emph{local} redundancy properties:

\begin{defi}\label{def:local}
A redundancy property $\mathcal{P}$ is \emph{local} if, 
for every clause $C$ and any two formulas $F,F'$ with $\env{F}(C) = \env{F'}(C)$, it holds that
$(F,C) \in \mathcal{P}$ if and only if $(F',C) \in \mathcal{P}$.
\end{defi}

\noindent
The following is easily seen to hold:

\begin{thm}
$\semblocked$ is a local redundancy property.
\end{thm}

\noindent Preparatory for showing
that $\semblocked$ is actually the most general 
local redundancy property
(cf.\ Theorem \ref{thm:semblocked_is_most_general} below), 
we first prove the following lemma:

\begin{lem}\label{thm:no_semblocked_no_locality}
If a clause $C$ is not semantically blocked in a formula $F$, 
then there exists a formula $F'$ with $\env{F'}(C) = \env{F}(C)$ such that 
$C$ is not redundant with respect to $F'$.
\end{lem}

\proof
Let $F$ be a formula and $C$ a clause that is not semantically blocked in $F$, i.e.,
there exists a satisfying assignment $\tau$ of $\envf(C)$ but there does not exist a satisfying assignment 
$\tau'$ of $\envf(C)  \cup \{C\}$ such that $\tau(v) = \tau'(v)$ for all $v \notin \var(C)$. 
We define the set $\tauliterals$ of (unit) clauses as follows:
\begin{align*}
	\tauliterals = \{(v) \mid \text{$v \notin \var(C)$ and $\tau(v) = \true$}\} \cup \{(\neg v) \mid \text{$v \notin \var(C)$ and $\tau(v) = \false$}\}\text{.}
\end{align*}
We further define the formula
$
	F' = \env{F}(C)
	\cup \{C\}
	\cup \tauliterals
	\text{.}
$

Since $C$ can be falsified (and is therefore not a tautology) 
and since the clauses in $\tauliterals$ contain only literals 
with variables that do not occur in $C$, it holds that
neither $C$ nor any of the clauses in $\tauliterals$ contain a literal $\bar{l}$ with $l \in C$. 
It therefore holds that $\env{F'}(C) = \env{F}(C)$.

Now observe the following: 
The assignment $\tau$ satisfies $\envf(C)$ and, by construction, also $\tauliterals$, hence 
$F' \setminus \{C\} = \envf(C) \cup \tauliterals$ is satisfiable. 
Furthermore, every satisfying assignment of $T$ 
must agree with $\tau$ on all variables $v \notin \var(C)$. 
But 
there exists no satisfying assignment $\tau'$ of $\envf(C)  \cup \{C\}$ 
such that $\tau(v) = \tau'(v)$ for all $v \notin \var(C)$.
Hence, $\envf(C)  \cup \{C\} \cup \tauliterals = F' \cup \{C\}$ is unsatisfiable 
and thus $F' \setminus \{C\}$ and $F' \cup \{C\}$ are not satisfiability equivalent.
It follows that $C$ is not redundant with respect to $F'$. 
\qed

\begin{thm}\label{thm:semblocked_is_most_general}
$\semblocked$ is the most general local redundancy property.
\end{thm}

\proof
Suppose there exists a local redundancy property $\mathcal{P}$ that is strictly more general than $\semblocked$. 
It follows that $\mathcal{P}$ contains a pair $(F,C)$ such that $C$ is not semantically blocked in $F$. 
By Lemma \ref{thm:no_semblocked_no_locality}, there exists a formula $F'$ with 
$\env{F'}(C) = \env{F}(C)$ such that $C$ is not redundant with respect to $F'$.
But since $\mathcal{P}$ is local and $\env{F'}(C) = \env{F}(C)$, 
it follows that \mbox{$(F',C) \in \mathcal{P}$}, hence $\mathcal{P}$ is not a redundancy property, a contradiction. 
\qed

\section{Set-Blocked Clauses and Super-Blocked Clauses}
\label{sec:super}

In the following, we introduce syntax-based notions of blocking which strictly generalize the original notion of literal-blocking 
as given in Definition \ref{def:bc_original}. 
We will first introduce the notion of set-blocking which is already a strict generalization of literal-blocking. 
This notion will then be further generalized to the so-called notion of super-blocking which, as we will prove, 
coincides with the notion of semantic blocking given in Definition \ref{def:semblocked}.

\begin{defi}\label{def:set_blocked}
Let $F$ be a formula and $C$ a clause. 
A non-empty set $L \subseteq C$ \emph{blocks} $C$ in $F$ if, 
for each clause $D \in F_{\bar L}$, 
$(C \setminus L) \cup \bar{L} \cup D$ is a tautology. 
We say that a clause is \emph{set-blocked in $F$} if there exists a set that blocks it in $F$.
We write $\setb$ to refer to $\{(F,C) \mid \text{$C$ is set-blocked in $F$}\}$.
\end{defi}

\begin{exa}\label{ex:setblocked}
Let $C = (a \lor b)$ and $F  = \{(\neg a \lor b), (\neg b \lor a)\}$. 
Then, $C$ is set-blocked by $L = \{a,b\}$: 
Clearly, $F_{\bar L} = F$ and $C \setminus L = \emptyset$. 
Therefore, for $D_1 = (\neg a \lor b)$ we get that $(C \setminus L) \cup \bar{L} \cup D_1 = (\neg a \lor b \lor \neg b)$ 
is a tautology and for $D_2 = (\neg b \lor a)$ we get that $(C \setminus L) \cup \bar{L} \cup D_2 = (a \lor \neg a \lor \neg b)$ is a tautology too. Note that $C$ is not literal-blocked in~$F$.
\end{exa}

\noindent Given an assignment $\tau$ that satisfies $F \setminus \{C\}$ but falsifies $C$, 
the existence of a blocking set $L$ guarantees that a satisfying assignment $\tau'$ of $F \cup \{C\}$ 
can be obtained from $\tau$ by flipping the truth values of all the literals in $L$. 
Since \mbox{$(C \setminus L) \cup \bar{L} \cup D$} is a tautology for every clause $D$ in the resolution environment of $C$, at least one of the following holds:
\begin{enumerate}[(i),ref={\roman*}]
	\item $D$ is itself a tautology and thus satisfied by $\tau'$, or
	\item $D$ contains a literal of $L$ which is satisfied by $\tau'$ since its truth value is flipped, or
	\item $D$ contains a literal $l$ which is satisfied since $\bar{l} \in C$ is falsified by $\tau$ and the truth value of $l$ is not flipped. 
\end{enumerate}
Hence, $\tau'$ satisfies $F \cup \{C\}$.

\begin{prop}\label{thm:bc_subset_setb}
Set-blocking is strictly more general than literal-blocking, i.e., it holds that $\bc \subset \setb$.
\end{prop}

\proof
Example \ref{ex:setblocked} shows that $\bc \neq \setb$.
It remains to show that $\bc \subseteq \setb$.
Let $F$ be a formula and let $C$ be a literal-blocked clause in $F$. 
We distinguish two cases:

\medskip\noindent
{\sc Case 1:} $C$ is a tautology. 
	We have that $\{l,\bar{l}\} \subseteq C$ for some literal $l$.
	Let $L = \{l,\bar{l}\}$.
	Then, $(C \setminus L) \cup \bar{L} \cup D$ is a tautology for every $D \in F_{\bar{L}}$.
		
\medskip\noindent
{\sc Case 2:} 
$C$ is not a tautology.
	Since $C$ is literal-blocked, there exists some literal $l \in C$ such that for every clause $D \in F_{\bar l}$, 
	$C \cup (D \setminus \{\bar{l}\})$ is a tautology.
	Let $L = \{l\}$ and let $D \in F_{\bar L}$.
	Then, since $F_{\bar L} = F_{\bar l}$, it follows that $C \cup (D \setminus \{\bar{l}\})$ is a tautology. 
	As $C$ is not a tautology, $D$ must contain some literal $l' \neq \bar{l}$ such that $\bar{l'} \in C \cup (D \setminus \{\bar{l}\})$. 
	Now, since $l' \neq \bar{l}$ we have that $\bar{l'} \neq l$ and thus $\bar{l'} \in (C \setminus \{l\}) \cup D$. 
	But then, $(C \setminus L) \cup \bar{L} \cup D$ is a tautology.

\medskip\noindent
It follows that $C$ is set-blocked in $F$ and therefore $\bc \subseteq \setb$.
\qed

\noindent We already argued informally why set-blocked clauses are redundant. 
That $\setb$ is indeed a redundancy property follows directly from the properties of super-blocked clauses, which we 
introduce next. 
To define super-blocked clauses, we first define the following formula modification that uses a (possibly partial) assignment to simplify a formula:

\begin{defi}
Given a formula $F$ and a assignment $\tau$, we denote by $F|\tau$ the set of clauses obtained from $F$ by removing all clauses that are satisfied by $\tau$.
\end{defi}

\noindent
Note that in the literature, $F|\tau$ sometimes denotes the set of clauses obtained from $F$ by first removing all clauses that are satisfied
by $\tau$ and then removing all literals that are falsified by $\tau$. For our purposes, it suffices to remove only satisfied clauses.
Before we next define super-blocked clauses, recall that the external variables, $\glob{F}(C)$, 
are the variables that occur in $\envf(C)$ but not in $C$.

\begin{defi}\label{def:superblocking}
A clause $C$ is \emph{super-blocked} in a formula $F$ if, for every assignment $\tau$ over the external variables 
$\glob{F}(C)$, $C$ is set-blocked in $F|\tau$.
We write $\superb$ for the set $\{(F,C) \mid \text{$C$ is super-blocked in $F$}\}$.
\end{defi}

\noindent Here, by ``{every assignment $\tau$ over the external variables $\glob{F}(C)$}'' we mean all assignments
whose domain is \emph{exactly} the set  $\glob{F}(C)$ and not a strict superset thereof. 
A simple example for a super-blocked clause is the clause $C$ in Example \ref{ex:full_blocking}---although $C$ is not set-blocked in~$F$, it is super-blocked in $F$ because it is set-blocked in both $F|\tau$ and $F|\tau'$ for $\tau(x) = 1$ and $\tau'(x) = 0$.
Again, the idea is that from an assignment $\tau$ that satisfies $F \setminus \{C\}$ but falsifies $C$, 
we can obtain a satisfying assignment $\tau'$ of $F \cup \{C\}$ by flipping the truth values of certain literals of $C$.
However, for making sure that the flipping does not falsify any clause $D$ in the resolution environment of $C$,
we are now also allowed to take into account the truth values of literals $l \in D$ with $\var(l) \in \glob{F}(C)$.
This is in contrast to set-blocking, where (apart from cases where $D$ is a tautology) we only consider
the truth values of literals whose variables occur in $C$.
Finally, note that if an assignment $\tau'$ is a superset of an assignment $\tau$, then $\tau'$ satisfies all clauses that
are satisfied by $\tau$ and thus the following statement holds:

\begin{prop}
Given a formula $F$ and two assignments $\tau, \tau'$, if $\tau \subseteq \tau'$, then $F|\tau' \subseteq F|\tau$.
\end{prop}

\noindent
From this it follows that if a clause is set-blocked in $F$, it is also set-blocked in $F|\tau$ for every assignment $\tau$. 
We conclude:

\begin{prop}
Super-blocking is strictly more general than set-blocking, i.e., it 
holds that $\setb \subset \superb$.
\end{prop}

\begin{thm}\label{thm:superblocked_iff_semblocked}
A clause is super-blocked in a formula $F$ if and only if it is semantically blocked in $F$, i.e., it holds that $\superb = \semblocked$.
\end{thm}

\proof
For the ``only if''
direction, let $C$ be a clause that is super-blocked in $F$ and let $\tau$ be a satisfying assignment of $\envf(C)$. 
If $\tau$ satisfies $C$, or $C$ contains a literal $l$ with $\var(l) \notin \var(F)$ 
(implying that $\tau$ can be straightforwardly extended to a satisfying assignment of $C$), 
then it trivially follows that $C$ is semantically blocked in $F$. 
Assume thus that $\var(C) \subseteq \var(F)$ and that $\tau$ does not satisfy $C$.
Furthermore, let $\tau_{E}$ be obtained from $\tau$ by restricting it to the external variables $\glob{F}(C)$. 
Since $C$ is super-blocked in $F$, there exists a non-empty set $L \subseteq C$ that blocks $C$ in $F|\tau_E$.
Consider the following assignment:
\begin{align*}
\tau'(v) = 
\begin{cases}
\false		& \text{if $\neg v \in L$},\\
\true			& \text{if $v \in L$},\\
\tau(v)		& \text{otherwise}.
\end{cases}
\end{align*}
Since $\tau$ falsifies $C$ 
there is no literal $l$ with $\{l, \bar l\} \subseteq L$,
hence $\tau'$ is well-defined. 
Clearly, $\tau'$ satisfies $C$ and $\tau'(v) = \tau(v)$ for all $v \notin \var(C)$. 
It remains to show that $\tau'$ satisfies $\envf(C)$. 
Since $\tau'$ and $\tau$ differ only on the truth values of variables in $\var(L)$, 
$\tau'$ can only falsify clauses containing a literal $\bar{l}$ with $l \in L$.
Let $D$ be such a clause. We proceed by a case distinction.

\medskip\noindent
{\sc Case 1:} $D$ contains an external literal $l$ (i.e., $\var(l) \in \glob{F}(C)$) that is satisfied by $\tau$. 
	Then, since $\var(l) \notin \var(C)$ and thus $l \notin L$, 
	it follows that $\tau'$ agrees with $\tau$ on the truth value of $l$, hence $l$ is satisfied by $\tau'$.	

\medskip\noindent
{\sc Case 2:}
$D$ does not contain an external literal that is satisfied by $\tau$.
	In this case, $D$ is contained in $F|\tau_E$ and thus, 
	since $L$ set-blocks $C$ in $F|\tau_E$, we have that $(C \setminus L) \cup \bar{L} \cup D$ is a tautology. 
	If $D$ is a tautology, then it is easily satisfied by $\tau'$, so assume that it is not a tautology. 
	Clearly, since $C$ is not a tautology, we have that $(C \setminus L) \cup \bar{L}$ is not a tautology, 
	hence there are two literals $l, \bar{l}$ such that $l \in D$ and $\bar{l}$ is in $C \setminus L$ or in $\bar{L}$. 
	If $\bar{l} \in C \setminus L$, then $\tau'$ agrees with $\tau$ on $\bar{l}$, 
	hence $\bar{l}$ is falsified by $\tau'$ and thus $l$ is satisfied by $\tau'$. 
	In contrast, if $\bar{l} \in \bar{L}$, then $l \in L$ and $l$ is satisfied by $\tau'$.
	In both cases
	$\tau'$ satisfies $l$ and thus it satisfies $D$.

\medskip
For the ``if'' direction, let $F$ be a formula and let $C$ be a clause that is not super-blocked in $F$, i.e., 
there exists an assignment $\tau_E$ over the external variables, $\glob{F}(C)$, 
such that $C$ is not set-blocked in $F|\tau_E$. Then, let
\begin{align*}
\tau(v) = 
\begin{cases}
\true		& \text{if $\neg v \in C$},\\
\false	& \text{if $v \in C$},\\
\tau_E(v)	& \text{otherwise}.
\end{cases}
\end{align*}
Clearly, $\tau$ is well-defined since $C$ cannot be a tautology, for otherwise it would be set-blocked in $F|\tau_E$. 
Furthermore, $\tau$ falsifies $C$ and, since (by definition) every clause $D \in \envf(C)$ 
contains a literal $\bar{l}$ such that $l \in C$, it satisfies $\envf(C)$.

Now let $\tau'$ be a satisfying assignment of $C$ such that $\tau'(v) = \tau(v)$ for all \mbox{$v \notin \var(C)$}. 
As $\tau'$ satisfies $C$, it is obtained from $\tau$ by flipping
the truth values of some literals $L \subseteq C$. 
We show that $\tau'$ does not satisfy $\envf(C)$. 
Clearly, $\tau'$ agrees with $\tau_E$ over the external variables, $\glob{F}(C)$, and since $C$ is not set-blocked in $F|\tau_E$, 
there exists a clause $D \in F|\tau_E$ with $D \cap \bar{L} \neq \emptyset$ such that 
$(C \setminus L) \cup \bar{L} \cup D$ is not a tautology and neither $\tau_E$ nor $\tau'$ satisfy any external literal in $D$.

Let now $l \in D$ be a (local) literal with $\var(l) \in \var(C)$. 
Since $(C \setminus L) \cup \bar{L} \cup D$ is not a tautology it follows that $\bar{l} \notin C \setminus L$ and $\bar{l} \notin \bar{L}$.
Since $\var(l) \in \var(C)$ we get that $l \in C \setminus L$ or $l \in \bar{L}$. 
In both cases, $l$ is not satisfied by $\tau'$. 
Thus, no literal in $D$ is satisfied by $\tau'$ and consequently
$\tau'$ does not satisfy $D \in \envf(C)$, which allows us
to conclude that $C$ is not semantically blocked in $F$.
\qed

\begin{cor}\label{thm:setb_is_a_local_redundancy_property}
$\setb$ is a \textsc{(}local\textsc{)} redundancy property.
\end{cor}

\section{The Relationship between Semantic Blocking and Variable Elimination}
\label{sec:ve}

In this section, we present an alternative characterization of semantic blocking based on 
Davis and Putnam's \emph{rule for eliminating atomic formulas}~\cite{davis60_a_computing_procedure}---a 
resolution-based rewriting rule which is also known as
\emph{variable elimination}~\cite{effective_preprocessing_een05} in the context of SAT solving.
Intuitively, a variable $x$ is eliminated from a formula by first adding all possible non-tautological resolvents 
upon $x$ and then removing all the clauses that contain $x$ or $\neg x$. 
In the following, for a literal~$l$, we denote by $F_l \res_l F_{\bar l}$ 
the set of all possible non-tautological resolvents upon~$l$, i.e., 
$F_l \res_l F_{\bar l} = \{C \res_l D \mid \text{$C \in F_l$ and $D \in F_{\bar l}$ and $C \res_l D$ is not a tautology}\}$.

\begin{defi}
Let $F$ be a formula, $x$ a variable, and $F' = F \setminus (F_x \cup F_{\bar x})$. 
The \emph{elimination of $x$ from $F$} is given by the formula $F' \cup (F_x \res_x F_{\bar x})$.
\end{defi}

\noindent
If $F$ does not contain any tautologies, then the variable $x$ does not occur in the resulting formula 
$F' \cup (F_x \res_x F_{\bar x})$. 
Moreover, variable elimination does not affect satisfiability, meaning that $F$ and $F' \cup (F_x \res_x F_{\bar x})$ are satisfiability equivalent~\cite{davis60_a_computing_procedure}.

\begin{exa}\label{ex:variable_elimination}
Let 
$F = \{(a \lor b), (x \lor b \lor \neg a), (\neg b \lor \neg x), (\neg b \lor a)\}$ 
\textup{(}cf.\ Example \ref{ex:full_blocking}\textup{)}. 
Then, 
$F_a = \{(a \lor b), (\neg b \lor a)\}$ and 
$F_ {\bar a} = \{(x \lor b \lor \neg a)\}$. 
Furthermore, 
$F' = F \setminus (F_a \cup F_{\bar a}) = \{(\neg b \lor \neg x)\}$ 
and $F_a \res_a F_{\bar a} = \{(b \lor x)\}$. 
Finally, the elimination of the variable $a$ from $F$ yields the formula 
$F_1 = F' \cup (F_a \res_a F_{\bar a}) = \{(\neg b \lor \neg x), (b \lor x)\}$.
\end{exa}

\noindent
In the rest of this section, we show that the following relationship between semantic blocking and variable elimination holds: 
To test whether a non-tautological clause is semantically blocked 
in a tautology-free formula, we can 
successively eliminate all the clause's variables from its resolution environment. 
If this elimination yields the empty formula, then the clause is semantically blocked, otherwise it is not.
The condition that the formula must be tautology-free is not a serious restriction as we can easily remove 
tautologies before variable elimination. 
%
%
The following example illustrates this relationship:

\begin{exa}
Consider again the formula $F$ from Example \ref{ex:variable_elimination}. 
As already shown earlier in Section~\ref{sec:super}, 
the clause $(a \lor b)$ from Example \ref{ex:variable_elimination} is super-blocked---and therefore semantically blocked---in~$F$.
Because of this, we would expect that 
the elimination of $a$ and $b$ from $F$ yields the empty formula. 
Indeed, in $F_1 = \{(\neg b \lor \neg x), (b \lor x)\}$ 
\textup{(}which, as shown in Example~ \ref{ex:variable_elimination}, 
is obtained from $F$ by eliminating the variable $a$\textup{)} there is only a tautological resolvent upon $b$, 
namely $(\neg x  \lor x)$, hence the elimination of $a$ and $b$ from $F$ yields the empty formula.
\end{exa}

\noindent
In order to prove that this relationship between variable elimination and semantic blocking holds, 
we first introduce a simple encoding of semantic blocking into quantified Boolean formulas (QBFs).
Given a clause $C$ and a propositional formula $F$, the encoding produces a quantified Boolean formula that is true 
if and only if $C$ is semantically blocked in~$F$.
Based on this encoding, we can then use a result from QBF theory that allows for a short proof of the main
statement of this section (Theorem~\ref{thm:empty_elimination_equals_superblocking} on page~\pageref{thm:empty_elimination_equals_superblocking}).
We therefore shortly recapitulate the syntax and semantics of quantified Boolean formulas~\cite{kleinebuening09_theoryofqbf}.

\begin{defi}
A \emph{quantified Boolean formula} (\mbox{QBF}) $\phi$ in 
\emph{prenex conjunctive normal form} (\mbox{PCNF}) is of the form $\quant . F$ 
where $\quant$ is a quantifier prefix and $F$, called the \emph{matrix} of $\phi$, is a propositional formula in CNF.
A \emph{quantifier prefix} has the form $Q_1 X_1 \dots Q_n X_n$
with disjoint variable sets $X_i$, $Q_i \in \{\forall, \exists\}$, and $Q_i \neq Q_{i+1}$.
\end{defi}

\noindent
A QBF $\forall x \quant.F$ is true if both $\quant.F[x/\top]$ and $\quant.F[x/\bot]$ are true, and false otherwise, where $\quant.F[x/t]$ is the
QBF obtained from $\quant.F$ by replacing all occurrences of the variable $x$ by the truth constant $t$. 
Moreover, a QBF $\exists x \quant.F$ is true if at least one of $\quant.F[x/\top]$ and $\quant.F[x/\bot]$ is true, and false otherwise.
If the matrix $F$ of a QBF $\phi$ is empty
after eliminating the truth constants
according to standard propositional rules, then $\phi$ is true. 
Accordingly, $\phi$ is false if $F$ contains the empty clause after eliminating the truth constants.
In the rest of this section, we assume every QBF to be in PCNF.

The following lemma introduces our encoding of semantic blocking into quantified Boolean formulas. 
Recall that for a clause $C$ and a formula $F$, $\glob{F}(C)$ denotes the set of external
variables of $C$, i.e., the set of variables that occur in the resolution environment of $C$ but not in $C$ itself.

\begin{lem}\label{thm:superblocked_iff_qbfsat}
Let $C$ be a non-tautological clause and $F$ a propositional formula. 
Let furthermore $G = \glob{F}(C)$ and $L = \var(C)$. 
Then, $C$ is semantically blocked in $F$ if and only if the {\normalfont QBF} \,$\forall G \exists L. (\envf(C) \cup \{C\})$ is true.
\end{lem}

\proof
For the ``only if'' direction, assume that $C$ is semantically blocked in $F$. 
We show that for every assignment $\tau_G$
over the variables in $G$, there exists an assignment $\tau_L$ over the variables in $L$ such that
$\tau_G \cup \tau_L$ satisfies $\envf(C) \cup \{C\}$.
Let $\tau_G$ be an arbitrary assignment over the variables in $G$.
To prove the existence of a corresponding $\tau_L$,
we first define the following assignment $\tau'_L$: 
\begin{align*}
\tau'_L(v) = 
\begin{cases}
\true		& \text{if $\neg v \in C$},\\
\false		& \text{if $v \in C$}.
\end{cases}
\end{align*}
Since $C$ is not a tautology, $\tau'_L$ is well-defined. 
Furthermore, since $\tau'_L$ falsifies all literals of $C$ and every clause $D$ in $\envf(C)$ 
contains a literal $\bar l$ with $l \in C$, the assignment $\tau_G \cup \tau'_L$ satisfies $\envf(C)$. 
Now, since $C$ is semantically blocked, there exists an assignment $\tau_{L}$ over the variables in $L$ 
such that $\tau_G \cup \tau_L$ satisfies $\envf(C) \cup \{C\}$. 
It therefore follows that the QBF $\forall G \exists L. (\envf(C) \cup \{C\})$ is true.

\medskip

For the ``if'' direction, assume that the QBF $\forall G \exists L. (\envf(C) \cup \{C\})$ is true 
and let $\tau$ be a satisfying assignment of $\envf(C)$.
We show that there exists a satisfying assignment $\tau'$ of $\envf(C) \cup \{C\}$
that agrees with $\tau$ on all the variables not occurring in~$C$.
To this end, let $\tau_G$ be obtained from $\tau$ by restricting it to the variables in $G$. 
Since $\forall G \exists L. (\envf(C) \cup \{C\})$ is true, 
there exists an assignment $\tau_L$ over the variables in $L$ such that 
$\tau' = \tau_G \cup \tau_L$ satisfies $\envf(C) \cup \{C\}$. As 
$\tau'$ agrees with $\tau$ on all the variables in~$G$, which are exactly the variables
that do not occur in $C$, it follows that $C$ is semantically blocked in $F$.
\qed

%

\bigskip
\noindent
With Lemma \ref{thm:superblocked_iff_qbfsat} we can give a short proof of 
Theorem \ref{thm:empty_elimination_equals_superblocking}
because it allows us to use an important result from QBF theory, 
namely that the elimination of existential variables from the inner-most 
quantifier block does not affect the truth value of a~QBF~\cite{heule14_a_unified_proof_system_for_qbf}:

\begin{lem}\label{thm:varelimination_qbf}
Let $\quant\exists X. F$ be a {\normalfont QBF} where $F$ does not contain any tautologies. Let furthermore $F'$ be obtained from $F$ by eliminating a variable~$x \in X$.
Then, $\quant\exists X. F$ is true if and only if $\quant\exists (X \setminus \{x\}).F'$ is true.
\end{lem}

\noindent
We can now finally state and prove the main result of this section:

\begin{thm}\label{thm:empty_elimination_equals_superblocking}
Let $C$ be a non-tautological clause, $F$ a propositional formula, 
and $E'$ obtained from $\envf(C) \cup \{C\}$ by
first removing all tautologies and then 
successively eliminating all variables that occur in $C$. 
Then, $C$ is semantically blocked in $F$ if and only if $E' = \emptyset$.
\end{thm}

\proof
Let $C$ be a non-tautological clause, $F$ a formula, 
and $E'$ obtained from $\envf(C) \cup \{C\}$ by
first removing all tautologies and then 
successively eliminating all variables that occur in $C$.
We show that $C$ is semantically blocked in $F$ if and only if $E' = \emptyset$.

Let $G = \glob{F}(C)$ and $L = \var(C)$. 
By Lemma \ref{thm:superblocked_iff_qbfsat}, $C$ is semantically blocked in $F$ if and only if the QBF
\begin{align*}
 \phi = \forall G \exists L. (\envf(C) \cup \{C\})
\end{align*}
is true. 
Now, since both the removal of tautologies and---by Lemma \ref{thm:varelimination_qbf}---the elimination of existential variables from the inner-most quantifier block of a QBF preserve its truth value, 
$C$ is semantically blocked in $F$ if and only if the QBF $\phi' = \forall G. E'$ is true, i.e., if the propositional formula $E'$ is valid.

For the ``only if'' direction, assume that $E' \neq \emptyset$, i.e., 
$E'$ contains a non-tautological clause $D$ such that $\var(D) \subseteq G$. 
Then, the assignment $\tau_G$ that is defined in such a way that it falsifies all literals of $D$, 
falsifies $E'$.
Hence, $\phi'$ is false and thus $C$ is not semantically blocked.

For the ``if'' direction, assume that $E' = \emptyset$. 
Then, $E'$ is trivially satisfied by every assignment over the variables in $G$ and thus $\phi'$ is true. 
It follows that $C$ is semantically blocked in $F$.
\qed

\noindent
We want to highlight that for Theorem~\ref{thm:empty_elimination_equals_superblocking}, the order in which variables are eliminated does not matter.
Theorem \ref{thm:empty_elimination_equals_superblocking} further clarifies the relationship between 
literal-blocking and semantic blocking: 
By definition, a non-tautological clause $C$ is literal-blocked in a formula $F$ if it contains a literal $l$ 
such that all resolvents upon~$l$ 
are tautologies. 
Since tautological resolvents are removed during variable elimination, 
we get the following alternative characterization of literal-blocking:


\begin{prop}\label{thm:empty_elimination_equals_blocking}
A non-tautological clause $C$ is blocked by a literal $l \in C$ in a formula $F$ 
if and only if the elimination of $\var(l)$ from $\envf(C) \cup \{C\}$ yields the empty formula.
\end{prop}

\noindent
In other words, literal-blocking requires that already the elimination of a single variable yields the 
empty formula while semantic blocking is more general by allowing for the elimination 
of several variables in order to derive the empty formula.

Note that Theorem \ref{thm:empty_elimination_equals_superblocking} does not hold 
without the condition that $C$ must be non-tautological.
To see this, consider the following example:

\begin{exa}
Let $C = (b \lor \neg b)$ and $F$ a formula in which $C$ has the resolution environment 
$\envf(C) = \{(a \lor \neg b), (\neg a \lor \neg b), (b \lor a), (b \lor \neg a)\}$. 
Then, $C$ is semantically blocked but the elimination of $b$ from $E' = \envf(C)$ \textsc{(}$C$ is not in $E'$ 
since it is a tautology\textsc{)} yields the \textsc{(}unsatisfiable\textsc{)} non-empty formula $\{(a), (\neg a)\}$.
\end{exa}

\noindent
To conclude this section, we note that the following theorem is a 
trivial consequence of Lemma~\ref{thm:superblocked_iff_qbfsat}:

\begin{thm}\label{thm:superblocked_characterization}
Let $C$ be a clause, $F$ a formula, $G = \glob{F}(C)$, and $L = \var(C)$. 
Then, $C$ is semantically blocked in $F$ if and only if it is a tautology 
or the {\normalfont QBF} $\forall G \exists L. (\envf(C) \cup \{C\})$ is true.
\end{thm}

\section{Complexity Analysis}
\label{sec:compl}

In this section, we analyze the complexity of testing whether a clause is set-blocked or super-blocked.
We further consider the complexity of testing restricted variants of set-blocking and super-blocking 
which gives rise to a whole family of blocking notions.
Note that all complexity results are with respect to the size of a clause and its resolution environment.

\begin{defi}
The \emph{set-blocking problem} is the following decision problem: 
Given a pair~$(F,C)$, where $F$ is a set of clauses and $C$ is a clause such that every clause $D \in F$ 
contains a literal $\bar{l}$ with $l \in C$, decide whether $C$ is set-blocked in $F$.
\end{defi}

\begin{thm}\label{thm:setblocking_is_npcomplete}
The set-blocking problem is \np-complete.
\end{thm}

\proof
We first show \np-membership, followed by 
\np-hardness. 

\medskip
\noindent
{\sc \np-membership}: 
For a non-empty set $L \subseteq C$, it can be checked in polynomial time whether 
$(C \setminus L) \cup \bar{L} \cup D$ is a tautology for every clause $D$ with $D \cap \bar{L} \neq \emptyset$. 
The following is thus an \np-procedure: 
Guess a non-empty set $L \subseteq C$ and check if it blocks $C$ in $F$.

\medskip
\noindent
{\sc \np-hardness}:
We give a reduction from \sat\ by
defining the following reduction function on input formula $F$ 
which is w.l.o.g.\ in CNF:
\begin{align*}
	f(F) = (F',C)\text{, with } C = (u \lor x_1 \lor x_1' \lor \dots \lor x_n \lor x_n')\text{,}
\end{align*}
where $\var(F) = \{x_1, \dots, x_n\}$ and $u, x'_1, \ldots, x'_n$ are new variables that do not occur in $F$. 
Furthermore, $F'$ is obtained from $F$ by
\begin{itemize}
	\item replacing every clause $D \in F$ by a clause $\ctransf(D)$, 
        obtained from $D$ by 
	adding $\neg u$ and replacing every negative literal $\neg x_i$ by the positive literal $x_i'$, 
	and
	
	\item adding the clauses	$(\neg x_i \lor \neg x_i'), (\neg x_i \lor u), (\neg x_i' \lor u)$ for every $x_i \in \var(F)$.
\end{itemize}
\noindent
The intuition behind the construction of $F'$ and $C$ is as follows.
By including $u$ in $C$ and adding $\neg u$ to every $\ctransf(D)$ 
with $D \in F$, we guarantee that all clauses in $F'$ are in the resolution environment of $C$, i.e., they contain a literal $l$ with $\bar{l} \in C$. 
This makes $(F',C)$ a valid instance of the set-blocking problem.
The main idea, however, is, that blocking-sets $L$ of $C$ in $F'$ correspond to satisfying assignments of $F$.
We show that $F$ is satisfiable if and only if $C$ is set-blocked in $F'$.

For the ``only-if'' direction, 
assume that there exists a satisfying assignment $\tau$ of $F$ 
and let $L = \{u\} \cup \{x_i \mid \tau(x_i) = \true\} \cup \{x_i' \mid \tau(x_i) = \false\}$. 
Clearly, $L \subseteq C$. 
It remains to show that for every $C' \in F'$ with $C' \cap \bar{L} \neq \emptyset$, 
$(C \setminus L) \cup \bar{L} \cup C'$ is a tautology. 
We proceed by a case distinction.

\medskip\noindent
{\sc Case 1:} $C'$ is of the form $(\neg x_i \lor u)$ or $(\neg x_i' \lor u)$ for $x_i \in X$. 
Then, since $\neg u \in \bar{L}$ and $u \in C'$, $(C \setminus L) \cup \bar{L} \cup C'$ is a tautology.

\medskip\noindent
{\sc Case 2:} $C'$ is of the form $(\neg x_i \lor \neg x_i')$ for $x_i \in X$. 
In this case, by the definition of $L$, only one of $x_i$ and $x_i'$ is in $L$. 
Assume w.l.o.g. that $x_i \in L$. 
Then, $x_i' \in C \setminus L$ and since $\neg x_i' \in C'$, $(C \setminus L) \cup \bar{L} \cup C'$ is a tautology.

\medskip\noindent
{\sc Case 3:} $C'$ is of the form $\ctransf(D)$ for $D \in F$. 
Then, $\tau$ satisfies a literal $l \in D$. If $l$ is positive, i.e., $l = x_i$ for some $x_i \in X$, 
then $x_i \in C'$ and $x_i \in L$. In contrast, if $l$ is negative, i.e., $l = \neg x_i$ for $x_i \in X$, 
then $x_i' \in C'$ and $x_i' \in L$. 
In both cases, $L$ contains a literal of $C'$. 
But then, $(C \setminus L) \cup \bar{L} \cup C'$ is a tautology.

\medskip
\noindent Thus, $L$ blocks $C$ in $F'$.

\medskip
For the ``if'' direction, suppose that $C$ is blocked by some blocking-set $L$ in $F'$ and define $\tau$ over $\var(F) = X$ as follows:
\begin{align*}
	\tau(x_i) = 
	\begin{cases}
	\true				& \text{if $x_i \in L$},\\
	\false			& \text{otherwise}.
	\end{cases}
\end{align*}
We show that $\tau$ satisfies $F$.
First, observe that $u$ must be contained in $L$: Assume to the contrary that $u \notin L$. 
Then, since $L$ is non-empty, some $x_i$ or $x_i'$ must be contained in $L$. If $x_i \in L$, 
then, since $C' = (\neg x_i \lor u)$ contains $\neg x_i$, $(C \setminus L) \cup \bar{L} \cup C'$ is a tautology. 
But, since $(C \setminus L) \cup \bar{L}$ cannot be a tautology, and $x_i \notin (C \setminus L) \cup \bar{L}$,
this can only be the case if $u \in L$, a contradiction. 
The argument for $x_i' \in L$ is analogous.

Now, let $D \in F$ and $C' = \ctransf(D)$. 
Then, since $u \in L$ and $\neg u \in C'$, $(C \setminus L) \cup \bar{L} \cup C'$ is a tautology. 
Furthermore, $C$ contains only positive literals and for $C'$ this is, apart from $\neg u$, also the case. 
But, since $u \in L$, $u$ is not contained in $(C \setminus L) \cup \bar{L}$ 
and thus $(C \setminus L) \cup \bar{L} \cup C'$ can only be a tautology 
if $C'$ contains a literal $l \in L$ which is different from $\neg u$. 
Now, if $l = x_i$ for some $x_i \in X$, then $x_i \in D$ and $\tau(x_i) = \true$. 
If $l = x_i'$ for $x_i \in X$, then $\neg x_i \in D$ and $\tau(x_i) = \false$.
In both cases, $\tau$ satisfies $D$. Hence, $F$ is satisfied by $\tau$.
\qed

\noindent 
We next analyze the complexity of testing whether a clause is super-blocked. 
To do so, we define the following problem:

\begin{defi}
The \emph{super-blocking problem} is the following decision problem:
Given a pair $(F,C)$, where $F$ is a set of clauses and $C$ a clause such that every $C' \in F$ 
contains a literal $\bar{l}$ with $l \in C$, decide whether $C$ is super-blocked in $F$.
\end{defi}

\begin{thm}\label{thm:semblocked_is_pi2complete}
	The super-blocking problem is \pitwo-complete.
\end{thm}

\proof
Since super-blocking coincides with semantic blocking, membership in \pitwo{} is an 
immediate consequence of Theorem \ref{thm:superblocked_characterization}. 
For showing hardness,
we give a reduction from \aesat{} to the super-blocking problem. 
The reduction is similar to the one used for proving Theorem \ref{thm:setblocking_is_npcomplete}.
Here, only the existentially quantified variables of the $\forall\exists$-formula are encoded into $C$ 
which makes all the universally quantified variables external.
Let $\phi = \forall X \exists Y F$ be an instance of \aesat{} with $Y = \{y_1,\dots,y_n\}$ and assume w.l.o.g.\ that $F$ is in CNF.
We define the reduction function
\begin{align*}
f(\phi) = (F',C)\text{, with } C = (u \lor y_{1} \lor y_1' \lor \dots \lor y_{n} \lor y_{n}')\text{,}
\end{align*}
where $u,y_1', \dots, y_n'$ are new variables not occurring in $\phi$. 
Furthermore, $F'$ is obtained from $F$ by
\begin{itemize}
	\item replacing every clause $D \in F$ by a clause $\ctransf(D)$ which is obtained from $D$ by 
	adding $\neg u$ and replacing every negative literal $\neg y_i$ by the positive literal $y_i'$ for $y_i \in Y$; and
	\item adding the clauses
	$(\neg y_i \lor \neg y_i'), (\neg y_i \lor u), (\neg y_i' \lor u)$ for every $y_i \in Y$.
\end{itemize}
By construction, every clause $C' \in F'$ contains a literal $\bar{l}$ with $l \in C$, 
hence $(F',C)$ is a valid instance of the super-blocking problem. 
The intuition behind the reduction is that blocking sets $L$ of $C$ in $F'$ 
correspond to assignments over the existential variables of $\phi$ 
while the assignments over the external variables, $\glob{F'}(C)$, 
correspond to the assignments over the universally quantified variables of $\phi$.
As super-blocking coincides with semantic blocking, 
we show that $\phi$ is true if and only if $C$ is semantically blocked in $F'$. 

For the ``only-if''
direction, assume that $\phi$ is true
and that there exists a satisfying assignment $\tau$ of $F'$.
We show that there exists a satisfying assignment $\tau'$ of $F' \cup \{C\}$ with $\tau'(v) = \tau(v)$ for all $v \notin C$.
Let therefore $\tau_{X}$ be obtained from $\tau$ by restricting it to the variables in $X$. 
Since $\phi$ is true, there exists an assignment $\tau_Y$ such that $\tau_{X} \cup \tau_Y$ satisfies $F$.
Consider now the following assignment:
\begin{align*}
	\tau'(v) = 
	\begin{cases}
	\tau_{X} \cup \tau_Y(v)	& \text{if $v \in X \cup Y$},\\
	\false				& \text{if $v = y_i'$ and $\tau_Y(y_i) = \true$ for $y_i \in Y$},\\
	\true					& \text{otherwise}.
	\end{cases}
\end{align*}
Clearly, $C$ is satisfied by $\tau'$ since $u \in C$ and $\tau'(u) = \true$ by definition.
Furthermore, $\tau'(v) = \tau(v)$ for all $v \notin C$ since all variables that are not in $C$ are contained in $\glob{F'}(C) = X$.
We show that $\tau'$ also satisfies $F'$.
Let therefore $C'$ be an arbitrary clause in $F'$. 
We proceed by a case distinction.

\medskip\noindent
{\sc Case 1:}
$C'$ is of the form $(\neg y_i \lor u)$ or $(\neg y_i' \lor u)$ for $y_i \in Y$. 
Then,	$C'$ is trivially satisfied.

\medskip\noindent
{\sc Case 2:}
$C'$ is of the form $(\neg y_i \lor \neg y_i')$ for $y_i \in Y$. 
Then, by definition of $\tau'$, $\tau'(y_i) \neq \tau'(y_i')$.
Hence, one of $y_i$ and $y_i'$ must be satisfied by $\tau'$.
	
\medskip\noindent
{\sc Case 3:}
$C' = \ctransf(C'')$ for $C'' \in F$.
	Since $\tau_{X} \cup \tau_Y$ satisfies $F$, there exists some literal $l \in C''$ such that $l$ is satisfied by $\tau_{X} \cup \tau_Y$. 
	Now, if $\var(l) \in X$ or $l$ is a positive literal, then $l$ is also contained in $C'$ and thus, 
$C'$ is satisfied by $\tau'$. 
	In contrast, if $l$ is a negative literal with $\var(l) \in Y$, 
	then
$C'$ contains the literal $y_i'$. 
	Since $\neg y_i$ is satisfied by $\tau_{X} \cup \tau_Y$ it follows that $\tau_Y(y_i) = \false$ and thus $\tau'(y_i') = \true$, 
	hence $C'$ is satisfied by $\tau'$.

\medskip\noindent
It follows that $\tau'$ satisfies $F' \cup \{C\}$.

\medskip
For the ``if''
direction, assume that $C$ is semantically blocked in 
$F'$ 
and let $\sigma_{X}$ be an arbitrary assignment over the variables in $X$.
We show that there exists an assignment $\sigma_{Y}$ over the variables in $Y$ such that $\sigma_{X} \cup \sigma_{Y}$ satisfies $F$.
To this end, we first define the following assignment over the variables in $F'$ and $C$:
\begin{align*}
\tau(v) = 
	\begin{cases}
	\sigma_{X}(v)	& \text{if $v \in X$},\\
	\false		& \text{otherwise}.
	\end{cases}
\end{align*}
Clearly, $\tau$ falsifies $C$ and since every $C' \in F'$ contains a literal $\bar{l}$ with $l \in C$, $\tau$ satisfies~$F'$.
Thus, since $C$ is semantically blocked in $F'$, there exists a satisfying assignment $\tau'$ of $F' \cup \{C\}$ 
such that $\tau'(v) = \tau(v)$ for all $v \notin \var(C)$. 
Since no variable from $X$ is contained in $C$, $\tau'$ agrees with $\sigma_{X}$ over the variables in $X$.
Now, let $\tau_{X \cup Y}$ be the assignment $\tau'$ restricted to the variables in $X \cup Y $. 
We show that $\tau_{X \cup Y}$ satisfies $F$.

Let $C'$ be an arbitrary clause in $F$.
By construction, $F'$ contains the clause $\ctransf(C')$ which is satisfied by $\tau'$.
Clearly, $\neg u \in \ctransf(C')$ is falsified by $\tau'$ because of the following:
The assignment $\tau'$ must satisfy some literal $l \in C$.
Since $l \in C$,
either $l = u$ or $F'$ contains the clause $(\bar{l} \lor u)$.
In both cases, since $\tau'$ satisfies $F' \cup \{C\}$,
$\tau'$ must satisfy $u$.
Hence, $\neg u$ is falsified by $\tau'$ and thus $\tau'$ must satisfy some literal $l'$ in $\ctransf(C')$ which is different from $\neg u$.
We proceed by a case distinction.

\medskip\noindent
{\sc Case 1:}
$\var(l') \in X \cup Y$.
	Then, by the definition of $\ctransf(C')$,
	$l' \in C'$ and since $\tau_{X \cup Y}$ agrees with $\tau'$ over $X \cup Y$,
	$l' \in C'$ is satisfied by $\tau_{X \cup Y}$.

\medskip\noindent
{\sc Case 2:}
$\var(l') \notin X \cup Y$.
	In this case, $l'$ is of the form $y'_i$.
	Since $\tau'$ satisfies $y'_i$ as well as the clause $(\neg y_i \lor \neg y'_i) \in F'$, it follows that $\tau'$ satisfies $\neg y_i$.
	Now, since $\ctransf(C')$ was obtained from $C'$ by adding $\neg u$ 
	and replacing every negative literal $\neg y_i$ by a literal $y_i'$, we get that 
	$\neg y_i \in C'$ is satisfied by $\tau'$ and since $\tau_{X \cup Y}$ agrees with $\tau'$ over $X \cup Y$,
	$C'$ is satisfied by $\tau_{X \cup Y}$.

\medskip\noindent
Hence, $\tau_{X \cup Y}$ satisfies $F$ and thus $\phi$ is true.
\qed

\noindent We have already seen that the set-blocking problem is \np-complete in the general case.
However, we obtain a restricted variant of set-blocking by only allowing blocking sets whose size is bounded by a constant.
Then, the resulting problem of testing whether a clause $C$ is blocked by some non-empty set $L \subseteq C$, 
whose size is at most $k$ for $k \in \posints$, turns out to be polynomial:
For a finite set $C$ and $k \in \posints$, there are only polynomially many non-empty subsets $L \subseteq C$ with $|L| \leq k$. 
To see this, observe (by basic combinatorics) that the exact number of such subsets is given by the following sum 
which reduces to a polynomial with degree at most $k$:
\[
\sum^{k}_{i=1}\binom{|C|}{i}\text{.}
\]
Hence, the number of non-empty subsets $L \subseteq C$ with $|L| \leq k$ is polynomial in the size of $C$. 
This line of argumentation is actually very common.
For the sake of completeness, however, we provide the following example:

\begin{exa}
Let $|C|=n$ and $k = 3$ \textup{(}with $k \leq n$\textup{)}. 
Then, the number of non-empty subsets $L \subseteq C$ with $|L| \leq k$ is given by the polynomial 
$$\sum^{3}_{i=1}\binom{n}{i} = \frac{n (n-1) (n-2)}{6} + \frac{n (n-1)}{2} + n = \frac{1}{6}n^3 + \frac{5}{6}n$$ of degree $k = 3$.
\end{exa}

\noindent
As there are only polynomially many potential blocking sets and since it can be checked in polynomial time 
whether a given set $L \subseteq C$ blocks $C$ in $F$
(as argued in the proof of Theorem \ref{thm:setblocking_is_npcomplete}),
it can be checked in polynomial time whether for some clause $C$ there exists a blocking set $L$ of size at most $k$.

Since the definition of super-blocking is based on the definition of set-blocking,
we can also consider the complexity of restricted versions of super-blocking 
where the size of the according blocking sets is bounded by a constant.
We thus define an infinite number of decision problems (one for every $k \in \posints$) as follows:

\begin{defi}
For any $k \in \posints$, the \emph{${k}$-super-blocking problem} is the following decision problem:
Given a pair $(F,C)$, where $F$ is a set of clauses and $C$ a clause such that every clause $D \in F$ contains a literal $\bar{l}$ with $l \in C$, 
decide if 
for every assignment $\tau$ over the external variables $\glob{F}(C)$, 
there exists a non-empty set $L \subseteq C$ with $|L| \leq k$ that blocks $C$ in $F|\tau$.
\end{defi}

\begin{thm}
The $k$-super-blocking problem is in \conp{} for all $k \in \posints$.
\end{thm}

\proof
\noindent Consider the statement that has to be tested for the complement of the $k$-super-blocking problem:

 \medskip
\begin{quote}
There exists an assignment $\tau$ over the external variables, $\glob{F}(C)$,
such that no non-empty subset of $C$ with $|C| \leq k$ blocks $C$ in $F|\tau$.
\end{quote}
\medskip
Since it can be checked in polynomial time whether a given set $L \subseteq C$ blocks $C$ in $F|\tau$, 
the following is an \np-procedure:

\medskip
\begin{quote}
Guess an assignment $\tau$ over the external variables, $\glob{F}(C)$, and check 
for every non-empty subset of $C$ (with $|C| \leq k$) whether it blocks $C$ in $F|\tau$.
If there is one, return \emph{no}, otherwise return \emph{yes}.
\end{quote}
\medskip
\noindent
Hence, for every integer $k \in \posints$, the $k$-super-blocking problem is in \conp.
\qed

\noindent Hardness for the complexity class \conp{} can be shown already for $k = 1$:

\begin{thm}
	The 1-super-blocking problem is \conp-hard.
\end{thm}

\proof
We show the result by providing a reduction from the unsatisfiability problem of propositional logic.
Let $F = \{C_1,\dots,C_n\}$ be a
formula in CNF
and define the reduction function
\begin{align*}
	f(F) = (F',C)\text{, with $C = (u_1 \lor \dots \lor u_n)$,}
\end{align*}
where $u_1,\dots,u_n$ are new variables that do not occur in $F$, 
and $F' = \bigcup_{i=1}^{n} F_i$ with $F_i = \{(\neg u_i \lor \bar{l}) \mid l \in C_i\}$.
Clearly, $(F',C)$ is a valid instance of the 1-super-blocking problem and $\var(F) = \glob{F'}(C)$.
We show that 
$F$ is unsatisfiable 
if and only if, 
for every assignment $\tau$ over $\glob{F'}(C)$, there exists a literal $u_i \in C$ such that $\{u_i\}$ set-blocks $C$ in $F'|\tau$.

For the ``only if'' direction, assume that $F$ is unsatisfiable and let $\tau$ be an assignment over $\glob{F'}(C)$. 
Since $\var(F) = \glob{F'}(C)$ it follows that there exists a clause $C_i$ in $F$ that is falsified by $\tau$. 
But then, since every clause in $F_i$ contains a literal $\bar{l}$ with $l \in C_i$, 
it follows that $F_i$ is satisfied by $\tau$. Hence, $F_i \cap F'|\tau = \emptyset$ and thus, 
since $\neg u_i$ only occurs in $F_i$, $\{u_i\}$ trivially set-blocks $C$ in $F'$.

For the ``if'' direction, assume that for every $\tau$ over $\glob{F'}(C)$, 
there exists a literal $u_i \in C$ such that $\{u_i\}$ set-blocks $C$ in $F'|\tau$. 
Since $\var(F) = \glob{F'}(C)$ it follows that for every assignment $\tau$ of $F$ and 
every clause $(\neg u_i \lor \bar{l}) \in F'|\tau$ (with $l \in C_i$), 
$T = (C \setminus \{u_i\}) \cup \{\neg u_i\} \cup \{\neg u_i, \bar{l}\}$ is a tautology. 
But since $T$ cannot contain complementary literals it must be the case that $(\neg u_i \lor \bar{l}) \notin F'|\tau$, 
which implies that every $l \in C_i$ is falsified by $\tau$. It follows that $F$ is unsatisfiable.
\qed

\noindent
The above reduction actually works for all $k$-super-blocking-problems with $k \in \posints$. 
To see this, observe that 
for every $k \in \posints$, $C$ is $k$-super-blocked in $F'$ 
if and only if 
it is 1-super-blocked in $F'$: 
If a clause is 1-super-blocked in a formula, then it is by definition also $k$-super-blocked for all $k \in \posints$. 
Conversely, due to the way $F'$ is constructed, if a set $L \subseteq C$ blocks $C$ in $F'|\tau$, 
with $\tau$ being an arbitrary assignment over $\glob{F'}(C)$, then there exists a singleton set 
$L' \subseteq L$ that blocks $C$ in $F'|\tau$ and thus $C$ is 1-super-blocked in $F'$. 
We thus get:

\begin{cor}
The $k$-super-blocking problem is \conp-complete for all $k \in \posints$.
\end{cor}

\noindent
The notions of set-blocking and super-blocking, together with the corresponding
restrictions discussed in this section, give rise to a whole family of blocking
notions which differ in both generality and complexity. We conclude
the following:
\begin{enumerate}[(i),ref={\roman*}]
	\item Considering the assignments over external variables (as is the case for super-blocking) leads to \conp-hardness.
	\item If blocking sets of arbitrary size are considered, the (sub-)problem of checking whether there exists a blocking set is \np-hard.
	\item If the size of blocking sets is bounded by a constant $k$, the \mbox{(sub-)}problem of testing whether 
	 there exists a blocking set turns out to be polynomial. 
	\item The problem of testing whether a clause is super-blocked in the most general sense, 
	where the size of blocking sets is not bounded by a constant, is \pitwo-complete.
\end{enumerate}
Hence, we can summarize the following complexity results:
\medskip
\begin{center}
\renewcommand{\arraystretch}{1.4}
\begin{tabular}{|l|c|c|}
	\hline
					& $|L|$ is unrestricted	&$|L| \leq k$ for $k \in \posints$\\
	\hline
	Super-Blocking 	& \pitwo-complete 		& \conp-complete\\
	\hline
	Set-Blocking		& \np-complete 		& \p\\
	\hline
\end{tabular}
\end{center}
\medskip
\noindent 
Note that the cardinality $|L|$ of blocking sets is of course bounded by the length of the 
clauses, thus we can restrict $|L| \leq |C|$.
This is particularly interesting for formula 
instances with (uniform) constant or maximal clause length.

Finally, we conclude the discussion by returning to the starting point 
of this paper: 
literal-blocked clauses.
Obviously, 
we can write the definition for set-blocking with $|L| \leq 1$ as follows:
A set $\{l\} \subseteq C$ \emph{blocks} a clause $C$ in a formula $F$ if 
for each clause $D \in F$ with $\bar{l} \in D$, $(C \setminus \{l\}) \cup D$ is a tautology. 
(Note that we write $(C \setminus \{l\}) \cup D$ instead of $(C \setminus \{l\}) \cup \{\bar{l}\} \cup D$ 
since $\bar{l}$ is anyhow required to be contained in $D$.) 
This is very similar to the original definition of literal-blocked clauses which requires $C \cup (D \setminus \{l\})$ to be a tautology.

\section{Comparison with Other Redundancy Properties}
\label{sec:comp}

In the following, we consider several local and non-local redundancy properties
as presented by J\"arvisalo et al.~\cite{Inprocessing} 
and relate them to the previously discussed local redundancy properties.
From the three basic redundancy properties of \emph{tautologies} ($\taut$), 
\emph{subsumed clauses}~($\subs$), and \emph{literal-blocked clauses} ($\bc$),  
extended redundancy properties are derived by \emph{asymmetric-literal addition} and/or a ``{resolution-look-ahead step}''.
We start with asymmetric-literal addition:

\begin{defi}
A literal $l$ is an \emph{asymmetric literal} with respect to a clause $C$ in a formula $F$ if $F \setminus \{C\}$ contains a clause
$D \lor \bar l$ such that $D \subseteq C$.
\end{defi}

\noindent
The addition of an asymmetric literal $l$ to a clause $C$ preserves equivalence in the sense that $F \setminus \{C\} \models (C \equiv C \lor l)$.

\begin{exa}
Consider the formula $F = \{(a \lor b), (b \lor c), (\neg c \lor d), (a \lor \neg c \lor \neg d)\}$ and let $C = (a \lor b)$.
The literal $\neg c$ is an asymmetric literal with respect to $C$ in $F$ because for the subclause $(b)$ of $(b \lor c)$ it holds that $(b) \subseteq (a \lor b)$.
Therefore, the addition of $\neg c$ to $C$ to preserves equivalence and we obtain the clause $C_1 = (a \lor b \lor \neg c)$. 
After this, $\neg d$ becomes an asymmetric literal with respect to $C_1$ because of $(\neg c \lor d)$. Adding $\neg d$ to
$C_1$ yields $C_2 =  (a \lor b \lor \neg c \lor \neg d)$. Now, $\neg a$ becomes an asymmetric literal with respect to
$C_2$ because of $(a \lor \neg c \lor \neg d)$ and so we add it to obtain the tautology $C_3 = (a \lor \neg a \lor b \lor \neg c \lor \neg d)$.
\end{exa}

\noindent
In the example above, the addition of asymmetric literals turned the clause $C$ into a tautology. 
Since tautologies are redundant and since asymmetric-literal addition preserves equivalence, 
we can infer that $C$ is redundant with respect to $F$. This leads to the notion of an asymmetric tautology:

\begin{defi}
A clause $C$ is an \emph{asymmetric tautology} in a formula $F$ if there exists a sequence $l_1, \dots, l_n$ 
of literals  such that $C \lor l_1 \lor \dots \lor l_n$ is a tautology and $l_i$ is an asymmetric literal with respect to $C \lor l_1 \lor \dots \lor l_{i-1}$ in $F$ for each $i \in 1, \dots, n$. 
By $\ataut$ we denote the redundancy property $\{(F,C) \mid \text{$C$ is an asymmetric tautology in $F$}\}$.
\end{defi}

\noindent
The notions of \emph{asymmetric subsumed clauses} ($\asubs$) and \emph{asymmetric blocked clauses} ($\abc$) are defined analogously:
A clause is an asymmetric subsumed clause if the addition of asymmetric literals turns it into a subsumed clause; 
it is an asymmetric blocked clause if the addition of asymmetric literals turns it into a blocked clause.

Asymmetric tautologies are particularly popular because they coincide with so-called $\rup$ clauses~($\rup$ stands for \emph{reverse unit propagation})~\cite{vangelder08_verifying_rup_proofs}. It follows for instance from results by Beame, Kautz, and Sabharwal~\cite{beame04_understanding_clause_learning} that the conflict clauses computed by conflict-driven-clause-learning SAT solvers are $\rup$ clauses, or, equivalently, asymmetric tautologies. Moreover, variants of asymmetric-literal addition and its converse, asymmetric-literal elimination, are used for minimizing clauses during SAT solving~\cite{wieringa13_concurrent_clause_strengthening,piette08_vivifying,heule11_efficient_cnf_simplification,luo17_learnt_clause_minimization,han07_alembic}. For instance, Luo et al.~\cite{luo17_learnt_clause_minimization} minimize a given learned clause $C$ by checking if there exists a subclause of $C$ that is an asymmetric tautology; if so, they replace the original clause by its \mbox{(stronger) subclause}.

Finally, we introduce redundancy properties that are intuitively obtained from existing ones
by performing an additional ``resolution-look-ahead step''.
Their names are obtained by adding the prefix $\reso$ to the abbreviated name of
the redundancy property they extend~\cite{Inprocessing}. More formally: 
Given a redundancy property $\mathcal{P}$, the pair 
$(F,C)$ is contained in $\reso\mathcal{P}$ if either
\begin{enumerate}[(i),ref={\roman*}]
	\item $(F,C) \in \mathcal{P}$, or 
	\item $C$ contains a literal $l$ such that for each clause $D \in F_{\bar l}$, $(F, C \cup (D \setminus \{\bar l\})) \in \mathcal{P}$. 
\end{enumerate}
Examples are $\rtaut$ (\emph{resolution tautologies}), $\rsubs$ (\emph{resolution-subsumed clauses}), and $\rat$ (\emph{resolution asymmetric tautologies}). Observe that resolution tautologies are nothing else than literal-blocked clauses.
Especially $\rat$~\cite{RAT, Inprocessing} is well-known: As almost all modern SAT solving techniques---including preprocessing, inprocessing, 
and clause learning---can be simulated by the addition and elimination of resolution asymmetric tautologies, they
provide the basis for the DRAT proof system~\cite{wetzler14_drattrim}, which is the standard in practical SAT solving.

The mentioned redundancy properties lead to the hierarchy depicted in Figure~\ref{fig:hier}.
We next show that our new redundancy properties are incomparable with
redundancy properties based on $\taut$;  
showing incomparability with subsumption-based properties works analogously.

\begin{figure}[t]
\centering
\usetikzlibrary{positioning}
\usetikzlibrary{decorations.markings}

\begin{tikzpicture}[node distance=0.5cm and 1cm]
\tikzstyle{local}=[rectangle,thick,draw=black,fill=black!10,minimum width=8mm,minimum height=5mm]
\tikzstyle{global}=[rectangle,thick,draw=black,fill=white,minimum width=8mm,minimum height=5mm]
\tikzstyle{new local}=[rectangle,thick,draw=black,fill=black!10,minimum width=8mm,minimum height=5mm]

\tikzstyle{arrow}=[thick, shorten >= 2pt, decoration={markings,mark=at position 1 with {\arrow[>=latex,thin]{>}}},
    postaction={decorate}]
\tikzstyle{bfarrow}=[thick, shorten >= 2pt, shorten <= 2pt, decoration={markings,mark=at position 0.14 with {\arrow[>=latex,thin]{<}}, mark=at position 1 with {\arrow[>=latex,thin]{>}}},
    postaction={decorate}]
\tikzstyle{new arrow}=[arrow]

\node [local] (t) {$\taut$};
\node [global,above= of t] (at) {$\ataut$};
\draw [arrow] (at) -- (t);
\node [global,left= of t] (s) {$\subs$};
\node [global,above= of s] (as) {$\asubs$};
\node [global,left= of s] (rs) {$\rsubs$};
\node [global,above= of rs] (ras) {$\rasubs$};
\node [local,right= of t] (b) {$\bc$};
\node [global,above= of b] (abc) {$\abc$};
\node [local,right= of b] (rt) {$\rtaut$};
\node [global,above= of rt] (rat) {$\rat$};
\node [new local,below= of b] (set) {$\setb$};
\node [new local,below= of rt] (sup) {$\superb$};

\node [local,minimum height=2mm,minimum width=2mm,yshift=0.28cm,xshift=-6.3cm,right= of set] (l){};
\node [right= of l,xshift=-0.63cm] {local};
\node [global,minimum height=2mm,minimum width=2mm,below= of l,yshift=0.3cm] (g){};
\node [right= of g,xshift=-0.6cm] {non-local};

\draw [arrow] (rs) -- (s);
\draw [arrow] (as) -- (s);
\draw [arrow] (ras) -- (rs);
\draw [arrow] (ras) -- (as);
\draw [arrow] (as) -- (at);
\draw [arrow] (abc) -- (at);
\draw [arrow] (rat) -- (abc);
\draw [arrow] (rat) -- (rt);
\draw [arrow] (rat) -- (rt);
\draw [bfarrow] (rt) -- (b);
\draw [arrow] (b) -- (t);
\draw [arrow] (abc) -- (b);
\draw [arrow] (rat.north) -- ([yshift=7pt]rat.north) -- ([yshift=7pt]ras.north) -- (ras.north);
\draw [new arrow] (set) -- (b);
\draw [new arrow] (sup) -- (set);

\end{tikzpicture}

\caption{
Hierarchy of redundancy properties~\cite{Inprocessing}
extended with novel local redundancies. 
An arrow from $\mathcal{P}_1$ to $\mathcal{P}_2$ denotes that 
$\mathcal{P}_1$ is more general than $\mathcal{P}_2$.
}
\label{fig:hier}
\end{figure}

\begin{prop}\label{thm:ht_and_bc_are_incomparable}
$\ataut \not\subseteq \setb$ and $\setb \not\subseteq \ataut$.
\end{prop}

%

\proof
	Let $C = (a \lor b \lor c)$ and $F = \{(\neg a \lor x), (\neg b \lor x), (\neg c \lor x), (a \lor b)\}$. 
	Because of the clause $(a \lor b)$, the literal $\neg b$ is an asymmetric literal with respect to $C$ and thus it follows that
	$(F,C) \in \ataut$.
	Now, assume that $C$ is set-blocked by some non-empty set $L \subseteq C$ in $F$, i.e.,
        for every clause $D \in F_{\bar L}$, 
        $(C \setminus L) \cup \bar{L} \cup D$ is a tautology. 
	Since $L$ is non-empty, at least one of
	$(\neg a \lor x)$, $(\neg b \lor x)$, and $(\neg c \lor x)$ must be contained in $F_{\bar L}$.
	Assume without loss of generality that $D = (\neg a \lor x)$ is contained in $F_{\bar L}$.
	Then, $a \in L$ and thus $a \notin C \setminus L$. Moreover, $a$ is not contained in $\bar L$.
	Hence, $(C \setminus L) \cup \bar{L} \cup D$ is not a tautology and thus $C$ is not set-blocked by $L$ in $F$,
	a contradiction.
	We conclude that $(F,C) \notin \setb$.
	
	To see that there exist set-blocked clauses that are not asymmetric tautologies, let $F = \emptyset$ and let $C = (a)$.
        Clearly, $(F,C) \in \setb$, but $(F,C) \not\in \ataut$.
\qed

\begin{prop}\label{thm:superblocked_and_hiddentaut_are_incomparable}
$\ataut \not\subseteq \superb$ and $\superb \not \subseteq \ataut$.
\end{prop}

\proof
	Consider again $C = (a \lor b \lor c)$ 
	and $F = \{$\mbox{$(\neg a \lor x)$}, $(\neg b \lor x), (\neg c \lor x), (a \lor b)\}$
	from the proof of Proposition \ref{thm:ht_and_bc_are_incomparable}, 
	and observe that $\glob{F}(C) = \{x\}$.
	Here, for the assignment $\tau$ that falsifies the external 
	variable $x$, 
	$F|\tau = F$ 
	and since $C$ is not set-blocked in $F$ (as shown in the proof of Proposition~\ref{thm:ht_and_bc_are_incomparable}), 
	it is not set-blocked in $F|\tau$, 
	hence $(F,C) \notin \superb$.
	
	To see that $\superb \not \subseteq \ataut$, let $F = \emptyset$ and $C = (a)$.
	Then, since $(F,C) \in \setb$ and \mbox{$\setb \subset \superb$}, 
	we get that 
	$(F,C) \in \superb$ but $(F,C) \notin \ataut$.
\qed

\noindent From Proposition \ref{thm:superblocked_and_hiddentaut_are_incomparable} 
together with the fact that $\ataut \subset \rat$ we get:

\begin{cor}
$\rat \not\subseteq \superb$.
\end{cor}

\begin{prop}
$\setb \not \subseteq \rat$.
\end{prop}

\proof
Consider the clause $C = (a \lor b)$ and the formula $F = \{(a \lor b),$ \mbox{$(\neg a \lor b)$}, \mbox{$(a \lor \neg b)$}$\}$. 
Clearly, $C$ is set-blocked by $L = \{a,b\}$ in $F$ and thus $(F,C) \in \setb$.

Now, for the literal $a$ there is only the clause $D_1 = (\neg a \lor b)$ that contains $\neg a$
and $C \cup D_1 \setminus \{\neg a\} = (a \lor b)$.
Furthermore, for the literal $b$ there is only the clause $D_2 = (a \lor \neg b)$ that contains $\neg b$ and 
here we again get that $C \cup D_2 \setminus \{\neg b\} = (a \lor b)$.
Since $(a \lor b)$ is not an asymmetric tautology in $F$, $(F,C) \notin \rat$.
\qed

\begin{cor}
$\rat$ is incomparable with both $\setb$ and $\superb$.
\end{cor}

\noindent
An intuitive explanation for the incomparability of $\setb$ and $\superb$ with both $\ataut$ and $\rat$ is the following: 
On the one hand, $\setb$ and $\superb$ have the advantage that they can flip the truth values of 
multiple literals when trying to determine a clause's redundancy, whereas $\ataut$ and $\rat$ can only flip the truth value of a single literal. 
On the other hand, $\setb$ and $\superb$ can only use information from the resolution environment 
of a clause while asysmmetric-literal addition allows $\ataut$ and $\rat$ 
to also use information from outside the resolution environment.

\section{Conclusion}
\label{sec:concl}

We showed that there exist redundancy properties that
are more general than blocked clauses while still being local, meaning that they can be checked 
by considering only the resolution
environment of a clause.
This locality aspect is part of the reason why
blocked clauses have been successful in the past
and it is particularly appealing
in the context of real-world verification
where problem encodings into SAT
often lead to very large formulas in which the resolution
environments of clauses are still small. 

By introducing a semantic notion of blocking, 
we characterized the essence of local redundancy: 
It suffices to flip only the truth values of some of the clause's literals to turn satisfying assignments of 
its resolution environment into assignments that satisfy also the clause itself.
With the aim of bringing this semantic blocking notion closer towards practical SAT solving,
we then introduced the syntax-based notions of set-blocking and super-blocking.
The notion of set-blocking strictly generalizes the traditional notion of blocking by allowing to flip the truth
values of more than only one literal.
Super-blocked clauses are even more general because the assignments 
on external variables are also taken into account when deciding redundancy. 
For super-blocked clauses, we proved that they coincide with
semantically blocked clauses.

In addition, we gave an alternative characterization 
of semantically blocked clauses based on variable elimination.
This characterization helps to clarify the relationship between traditionally blocked clauses
and semantically blocked clauses: 
Traditional blocking requires that 
already the elimination of a single variable from a clause and its resolution environment yields the empty formula. 
In comparison, semantic blocking is more general as it allows for the elimination of 
several variables in order to derive the empty formula.
Our complexity analysis showed that checking the newly introduced redundancy properties 
is computationally expensive in the worst case.
At first glance, this seems to limit their practical applicability.
However, we presented bounded variants that can be checked more efficiently
and we expect them to improve the solving process considerably when added to our SAT solvers.

While the focus of this paper lies on the theoretical investigation of local redundancy properties, thereby
contributing to gaining a deeper understanding of blocked clauses,
a practical evaluation is subject to future work.
Another direction for future work is lifting the new redundancy properties 
to QSAT, the satisfiability problem of quantified Boolean formulas (QBF). 
There, blocked clauses have shown to be practically even more effective than in 
SAT solving~\cite{ClauseEliminationJournal,DynamicBlockedClauses}
and we expect this to also hold for quantified variants of 
set-blocked clauses and super-blocked clauses.

\bibliography{references}
\bibliographystyle{plain}

\end{document}